\documentclass{ws-procs961x669}     
 \usepackage{graphicx} 
 \newcommand{\LF}{\left(}
 \newcommand{\RF}{\right)}
 \newcommand{\LT}{\left[}
 \newcommand{\RT}{\right]}
 \pdfpageheight\paperheight
 \pdfpagewidth\paperwidth

 \def\Dc{\mathcal{D}}
 
 \def\Fc{\mathcal{F}}
 
 \def\Hc{\mathcal{H}}

 \def\Kc{\mathcal{K}}
 \def\Lc{\mathcal{L}}
 \def\Mc{\mathcal{M}}
 
 \def\Oc{\mathcal{O}}
 \def\Pc{\mathcal{P}}
 
 \def\Rc{\mathcal{R}}

 \def\Wc{\mathcal{W}}

\begin{document}
\title{Non-local $R^2$-like inflation, Gravitational Waves and Non-Gaussianities}

\author{K. Sravan Kumar$^*$}

\address{Department of Physics, Tokyo Institute of Technology\\
1-12-1 Ookayama, Meguro-ku, Tokyo
152-8551, Japan\\
$^*$E-mail: sravan.k.aa@m.titech.ac.jp,\\
www.titech.ac.jp/english}

\begin{abstract}
The emergence of $R^2$ (Starobinsky) inflation from the semi-classical modification of gravity due to matter quantum fields (trace anomaly) clearly points out the importance of fundamental physics and the first principles in the construction of successful cosmological models. Along with the observational success, $R^2$ gravity is also an important step beyond general relativity (GR) towards quantum gravity. Furthermore, several approaches of quantum gravity to date are strongly indicating the presence of non-locality at small time and length scales. In this regard, ultraviolet (UV) completion of $R^2$ inflation has been recently studied in a string theory-inspired ghost-free analytic non-local gravity. We discuss the promising theoretical predictions of non-local $R^2$-like inflation with respect to the key observables such as tensor-to-scalar ratio, tensor tilt which tell us about the spectrum of primordial gravitational waves, and scalar Non-Gaussianities which tell us about the three-point correlations in the CMB fluctuations. Any signature of non-local physics in the early Universe will significantly improve our understanding of fundamental physics at UV energy scales and quantum gravity.
\end{abstract}


\bodymatter

\section{Status of inflation and the quests for UV-completion}\label{aba:sec1}
Inflationary cosmology has become the most important field of study in theoretical higher energy physics (THEP) in the recent years not only because of its success with respect to the data of Cosmic Microwave Background (CMB) \cite{Akrami:2018odb,BICEP:2021xfz,Akrami:2019izv} but also it gives a best chance to test our approaches to build physics at the fundamental scales such as Planck energy or length scales. Recent detection of Gravitational Waves from binary black hole merger events further corroborated to build new hopes and strategies to probe the physics of early Universe \cite{Abbott:2016blz,Ricciardone:2016ddg}.  

\begin{figure}[h!]
	\centering
	\includegraphics[width=0.45\linewidth]{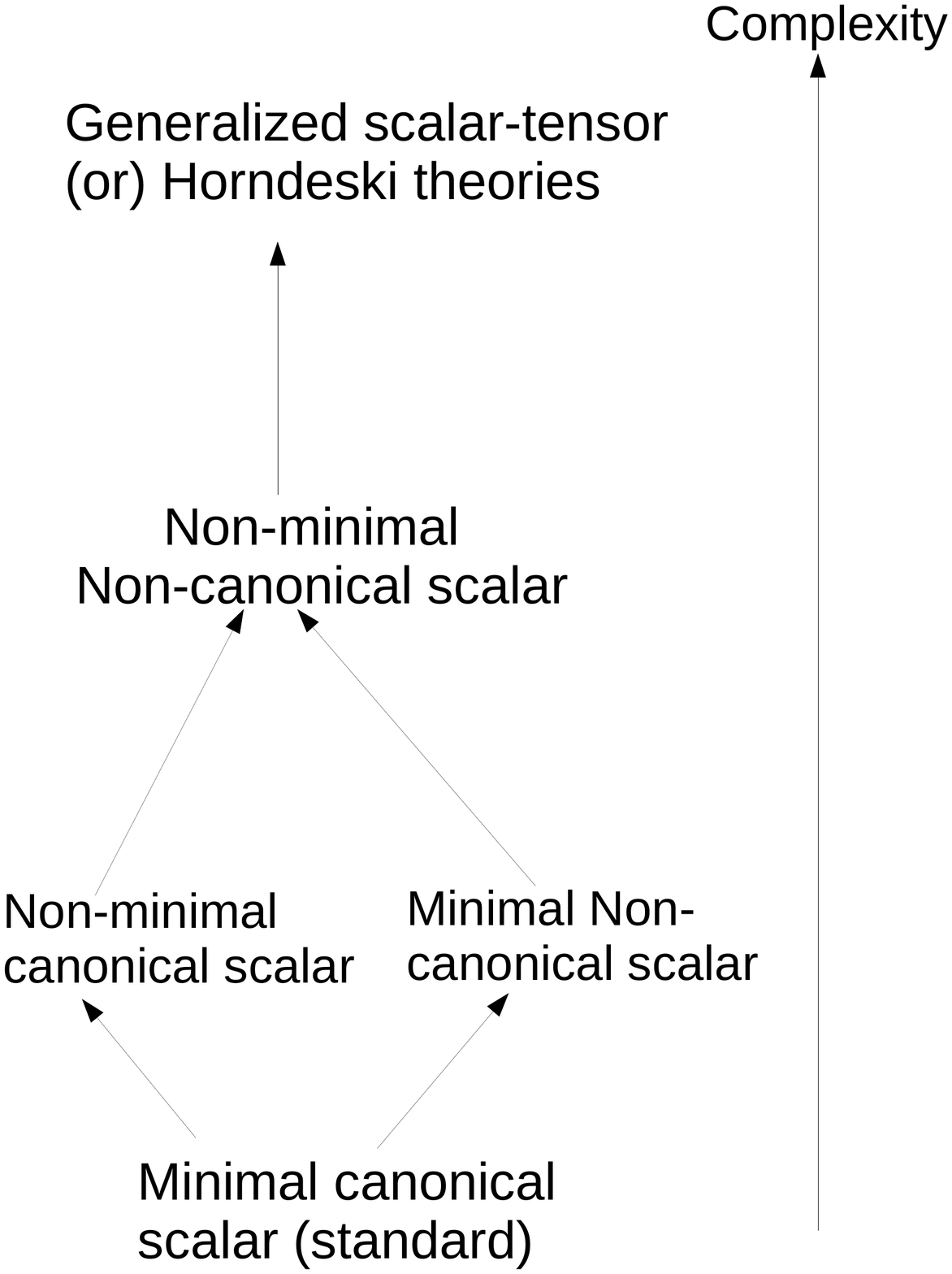}\quad \includegraphics[width=0.45\linewidth]{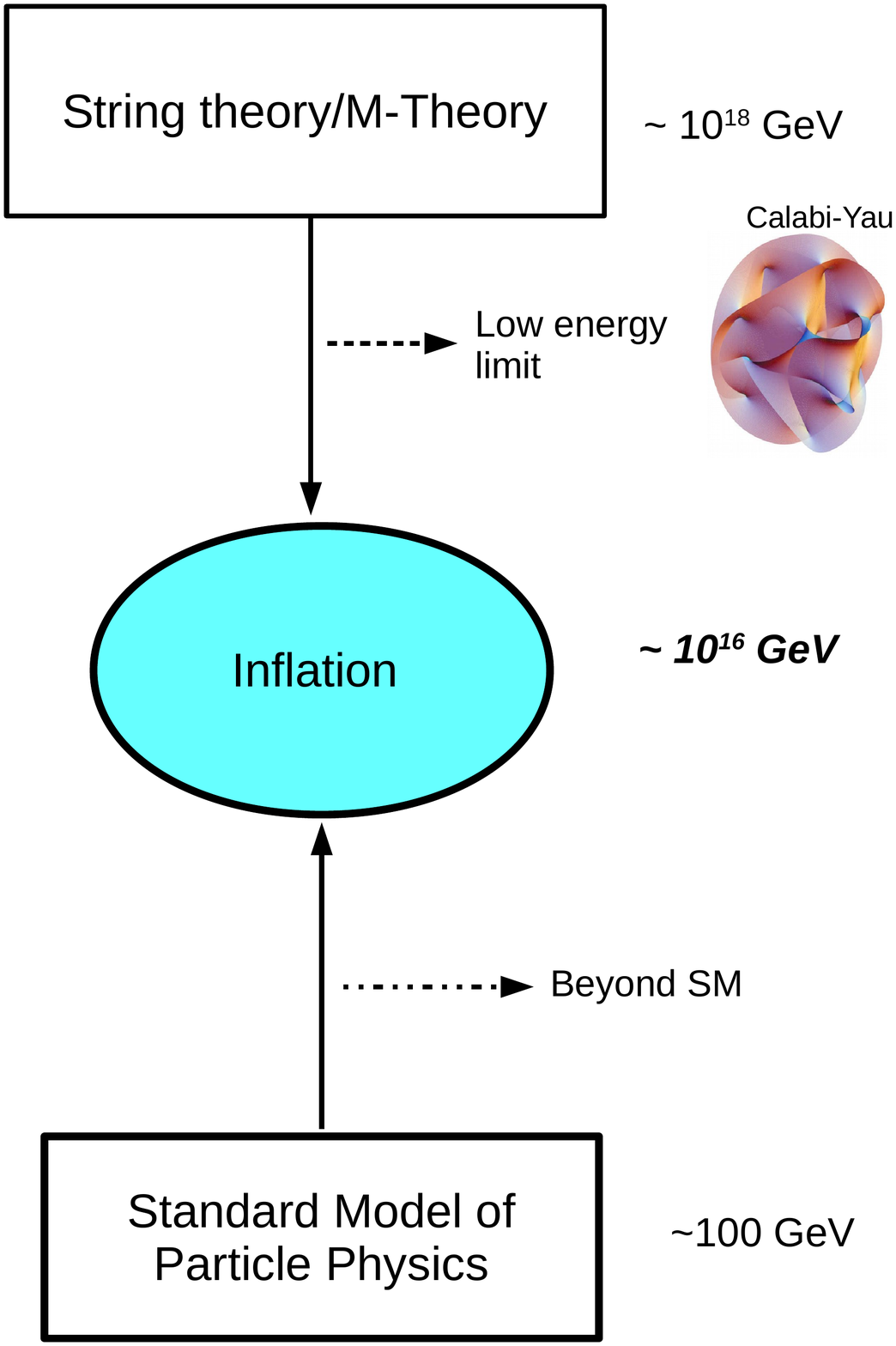}
	\caption{In the left panel, arrow of complexity shows how one can start with a simplest scalar field Lagrangian and generalized all the way up to most general scalar-tensor theories whose equations of motion are second order in field derivative. In the right panel we can see that inflationary cosmology can be understood well either from beyond the standard model of particle physics or from low-energy limit of a UV-complete approaches such as string theory (This Figure taken from \cite{SravanKumar:2018reb}).}
	\label{fig:draw1}
\end{figure}

Inflationary cosmology is so far developed in two equivalent ways. One is by geometrical modification i.e., by extension of General Relativity (GR) by higher curvature terms (for example "$R+R^2$ gravity which is also known as Starobinsky theory \cite{Starobinsky:1980te,Starobinsky:1981vz,Starobinsky:1981zc,Starobinsky:1984vz}). Second way is by the addition of hypothetical (scalar) matter fields which result in modification of right hand side of Einstein equations \cite{Guth:1980zm,Linde:1983gd}.  This second way motivated in the view of particle physics beyond the standard model and several ultraviolet (UV) complete approaches such as string theory and supergravity (SUGRA) \cite{Linde:2014nna}. In this regard, inflationary cosmology is so far studied from the point of view of beyond the standard model of particle physics and as a low-energy limit of UV-complete approaches such as string theory/M-theory (see right panel of Fig.~\ref{fig:draw1}). In the last decades plethora of inflationary models have been constructed with different scalar fields emerging from simplest set ups to all the way up to the most general scalar tensor theories \cite{SravanKumar:2018reb,Martin:2013tda} (see left panel of Fig.~\ref{fig:draw1}).  However the latest data from Planck satellite \cite{Akrami:2018odb} has found out Starobinsky and Higgs inflation be the most compatible models with the following predictions of spectral index and the tensor to scalar ratio 
\begin{equation}
	n_{s}=1-\frac{2}{N}\,,\qquad r=\frac{12}{N^{2}}\,, 
\end{equation}
where $N=50-60$ number of e-folds before the end of inflation. The recent Planck data \cite{Akrami:2018odb,BICEP:2021xfz} has constrained $ n_s= 0.9649\pm 0.0042\,\, \textrm{at}\,\, 68\% \textrm{CL}, r<0.036 \,\,\textrm{at}\,\, 95\%\, \textrm{CL}$. 
The success of Starobinsky and Higgs inflation has lead to the emergence of  equivalent frameworks of inflation in string theory and SUGRA \cite{Linde:2014nna}. 

The challenging problem of the current status of inflationary cosmology is to break several degeneracies in the frameworks of inflation and find definitive clues towards achieving quantum gravity and UV-completion. For this purpose, from observational point of view, one must go beyond the two observables $(n_s,\,r)$. In this regard there are two more important observational parameters known as tensor tilt ($n_t$) which is the tilt of the tensor power-spectrum and non-Gaussianity parameter ($f_{NL}$) which is related to the 3-point scalar correlations. If inflation is driven by single (canonical) scalar field with standard slow-roll then the tensor consistency relation \cite{SravanKumar:2018reb} and the Maldacena consistency relation \cite{Maldacena:2002vr} given by
\begin{equation}
	r= -8n_t \quad f^{\rm sq}_{NL} = \frac{5}{12} (1-n_s)
	\label{consis}
\end{equation}
serves as a crucial test. Here $f^{\rm sq}_{NL}$ is so-called squeezed shape of non-Gaussianity. If inflation is driven by multiple fields then the above consistency relations are understood to be violated \cite{Meerburg:2019qqi}. If inflation is driven by non-canonical scalar fields and if non-standard quantum initial conditions are assumed we will find a strong signal in the equilateral ($f^{\rm eq}_{NL}$) and orthogonal ($f^{\rm ortho}_{NL}$) shapes of non-Gaussianities. These shapes are constrained by the latest Planck data as 
\begin{equation}
		f_{\rm NL}^{\rm loc}=0.8\pm 5.0 \,,\quad f_{\rm NL}^{\rm equi}=-4\pm 43 \,,
	\quad f_{\rm NL}^{\rm ortho}=-26\pm 21\,,\quad 
	{\rm at} \,\, \,68\% {\rm CL}.
\end{equation}
Violation of consistency relations \eqref{consis} and any detection of $f_{NL}\sim O(1)$ is definitely a new physics in the context of early Universe cosmology which will be crucial for further advancements in THEP. Moreover, in a quest to find signatures of new physics, recently a new program of research has gained significant attention known as "cosmological collider physics" \cite{Arkani-Hamed:2015bza} whose aim is to find signatures of higher energy particles interacting with primordial fields during inflation. 

Despite numerous phenomenological approaches in understanding inflation and find new observables to probe physics of inflation, it is very vital to focus on consistency of our theoretical endeavors. The success of Starobinsky and Higgs inflationaris (which are almost identical during inflation) covey us a very crucial property that at the scale of inflation approximate  "scale invariance" symmetry of the Lagrangian is important. The fact that various setups of SUGRA leading to Starobinsky-like or $R^2$-like inflation \cite{Linde:2014nna} indicate that $R+R^2$ is an important modification of GR. In the next section, we will discuss in detail about the foundations of Starobinsky inflation and how a very straightforward procedure from geoemertric modification of gravity lead us towards a UV-complete framework \cite{Koshelev:2016xqb,Koshelev:2017tvv,Koshelev:2020foq,Koshelev:2020xby}. 

\section{From Starobinsky inflation to Stelle gravity and the emergence of ghost-free UV-non-local gravity}

$R+R^2$ gravity should not be viewed as a toy model in the class of f(R) gravity. The theory is well-motivated from generic considerations and important questions about UV-completion which dates back the discovery of trace anomaly \cite{Duff:1993wm}. The trace anomaly comes from 1-loop quantum corrections to the graviton propagator due to matter conformal fields leading to an anomalous non-zero trace of energy momentum tensor which is expressed as
\begin{equation}
	M_p^2 R = g^{\mu\nu}\langle T_{\mu\nu} \rangle = a F+ bG+c\square R\,,  
	\label{tra}
\end{equation} 
where $F=W_{\mu\nu\rho\sigma}W^{\mu\nu\rho\sigma}$ is the 
the Weyl tensor square and $G= R_{\mu\nu\rho\sigma}R^{\mu\nu\rho\sigma}-4 R_{\mu\nu}R^{\mu\nu} +R^2$ is the so-called Gauss-Bonnet term and  $R_{\mu\nu\rho\sigma}$, $R_{\mu\nu}$, $R$ Riemann tensor, Ricci tensor and Ricci scalar respectively. Here $a,b,c$ are dimensionless coefficients quantified by the number of conformal fields. Trace anomaly has been widely studied and applied in various fundamental theories including anticipated UV complete theories such as string theory and supergravity (SUGRA) \cite{Duff:1993wm}. 
Neglecting the coefficients $a, b$ the trace anomaly equation reduces to the trace equation of the $R+R^2$ given by the action \cite{Starobinsky:1980te,Hawking:2000bb} 
\begin{equation}
	S_{R^2}= \int d^4x\sqrt{-g} \left[ \frac{M_p^2}{2}R + \frac{M_p^2}{12M^2} R^2 \right]\,. 
	\label{LSR2}
\end{equation}
where $M\ll M_p$ becomes the mass of a propagating scalar in this model, named scalaron. Independent of what is the right framework of UV-completion  $R^2$ term is very natural term that appears as a classical addition to the GR and it naturally leads to a quasi-de Sitter evolution (which is "cosmic inflation") in the high  curvature regime $R\gg M^2$. Therefore it is a matter of fact that $R+R^2$ gravity indeed an unavoidable and very important extension of GR. Furthermore, if we add Weyl square term to the action of $R+R^2$ gravity \eqref{LSR2} the resultant gravity theory becomes renormalizable (which is an important property to achieve quantum gravity) as it was proven by K. S. Stelle long ago \cite{Stelle:1976gc} 
\begin{equation}
	S_{\rm Stelle} =  \int d^4x\sqrt{-g} \left[ \frac{M_p^2}{2}R + \frac{M_p^2}{12M^2} R^2 +\frac{M_p^2}{M_g^2}  W_{\mu\nu\rho\sigma}W^{\mu\nu\rho\sigma}  \right]
\end{equation}
However, Stelle gravity unfortunately falls short in becoming a consistent theory of quantum gravity because of the presence of a massive tensor ghost in the spectrum which spoils Unitarity. A natural question one can ask is that whether this ghost can be removed by any higher derivative extension of Stelle gravity. Due to Ostrogradsky theorem any finite derivative extension of Stelle gravity results in further introduction of ghosts in the spectrum and it becomes not a viable option. However, Krasnikov has shown that analytic infinite derivative extension of Stelle gravity (i.e., analytic non-local gravity) can give  a ghost-free theory \cite{Krasnikov:1987yj}.

\begin{equation}\label{NC-action}
	S=\int d^{4}x\sqrt{-g}\left(\frac{M_{p}^{2}}{2}R+%S_{1},\label{NC-action}
	\frac{1}{2}
	\bigg[R\mathcal{F}_{R}\left(\square_{s}\right)R+W_{\mu\nu\rho\sigma}\mathcal{F}_{W}\left(\square_{s}\right)W^{\mu\nu\rho\sigma}\bigg]\right)\,.%\label{NCA}
\end{equation}
where $\square_s=\frac{\square}{\Mc_s^2}$ with $\square$ is being the d'Alembertian and  $\Fc_R, \Fc_W$ are analytic infinite derivative operators or non-local form factors.  This theory \eqref{NC-action} has been studied over the years and understood to have good properties of UV-completion with respect to renormalizability and Unitarity  (See \cite{Koshelev:2020xby} and the references therein). In a nutshell, the action \eqref{NC-action} is a very logical extension of Stelle gravity and we very much straightforwardly end up with introducing non-locality in the gravity sector. Note that "Non-locality" is a very important common concept that appears in several approaches to quantum gravity such as string theory, loop quantum gravity, causal sets and non-commutative gravity \cite{Koshelev:2020xby,Buoninfante:2021xxl}. Therefore, \eqref{NC-action} is a consistent culmination of various approaches which allow us to give a consistent UV-completion for Starobinsky inflation\footnote{Worth to metion that Starobinsky inflation has also been embedded in SUGRA which gives an interesting route for UV-completion \cite{Ketov:2010qz}. Here we focus on non-supersymmetric aspects of UV-completion.}. Further note that \eqref{NC-action} should not be understood as an effective theory but rather a consistent part of a UV-complete or a fundamental theory. We do not have any cut off scales in the theory and the scale $\Mc_s$ represents the scale at which non-local nature of gravity is prominent. In the next section, we will discuss the cosmological application of \eqref{NC-action}  and identify the specific features of non-local gravity which can be probed in the future observations. 

\section{Non-local $R^2$-like inflation and future targets for CMB and Gravitatinal wave observations} 

\begin{figure}
	\centering
	\includegraphics[width=0.7\linewidth]{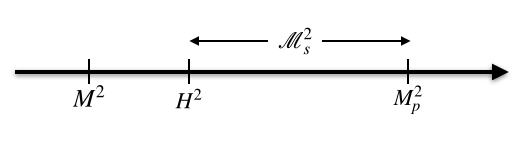}
	\caption{Hierarchy of energy scales in non-local $R^2$-like inflation.}
	\label{fig:scales}
\end{figure}

In this section, we apply UV non-local quadratic curvature gravity in the context of inflationary cosmology. The first and foremost difficulty is associated with solving equations of motion of an infinite derivative theory which perhaps appears to be an impossible task. However, this was consistency achieved \cite{Koshelev:2016xqb} and it was also quite rigorously shown in \cite{Koshelev:2017tvv} that the theory contains an inflationary attractor solution exactly same as the local $R+R^2$ or Starobinsky theory. To see this explicitly let us write down the background equations of motion for \eqref{NC-action} Friendmann-Lema\^itre-Robertson-Walker background for which the Weyl tensor is zero 
	\begin{equation*}
	\begin{aligned}\bar{E}_{\nu}^{\mu}\equiv & -\left[M_{p}^{2}+2\lambda\Fc\left(\frac{\square}{\Mc_s^2}\right)\bar{R}\right]\bar{G}_{\nu}^{\mu} \delta^{\mu}_{\nu}-\frac{1}{2}\bar{R}\Fc\left(\frac{\square}{\Mc_s^2}\right)\bar{R}\delta_{\nu}^{\mu}\\
		&+2\left(\nabla^{\mu}\partial_{\nu}-\delta_{\nu}^{\mu}\square\right)\Fc\left(\frac{\square}{\Mc_s^2}\right)R +\mathcal{\bar{K}}_{\nu}^{\mu}-\frac{\lambda}{2}\delta_{\nu}^{\mu}\left(\mathcal{\bar{\Kc}}_{\sigma}^{\sigma}+\bar{\mathcal{K}}\right)=0\,,
	\end{aligned}
	\label{EoM}
\end{equation*}
where

\[
\begin{aligned}\mathcal{\bar{\Kc}}_{\nu}^{\mu}= \frac{1}{\Mc_s^2} \sum_{n=1}^{\infty}f_{n}\sum_{l=0}^{n-1}\partial^{\mu}\frac{\square^{l}}{\Mc_s^{2l}}\bar{R}\partial_{\nu}\left(\frac{\square}{\Mc_s^2}\right)^{n-l-1}\bar{R}\,,
	\bar{\mathcal{K}}=  \sum_{n=1}^{\infty}f_{n}\sum_{l=0}^{n-1}\frac{\square^{l}}{\Mc_s^{2l}}\bar{R}\left(\frac{\square}{\Mc_s^2}\right)^{n-l}\bar{R}\,.
\end{aligned}
\]
The trace equation is  
\begin{equation} 
	\bar{E}=M_p^2 \bar{R}-6\bar{\square}\Fc\left(\frac{\bar{\square}}{\Mc_s^2}\right) \bar{R}-\bar{\Kc}_\mu^\mu-2\bar{\Kc}=0 \, .
	\label{tEOMtrace}
\end{equation}
The overbars in \eqref{EoM} and \eqref{tEOMtrace} is used to indicate background quantities. 
It was showed in \cite{Koshelev:2017tvv} that the only solution of the above equations is $\bar{\square}\bar{R} = M^2\bar{R}$ given the following Unique conditions on the form factor are satisfied. 
\begin{equation}
	\Fc^{\prime}_R\LF\frac{M^2}{M_s^2}\RF=0,~\frac{M_{P}^{2}}{2\lambda}=3M^2\mathcal{F}_{1},~\text{ where }\Fc_1\equiv\Fc_R\left(\frac{M^2}{\Mc_s^2}\right)
	\label{conditions}
\end{equation} 
where $^\prime$ denotes derivative with respect to argument. 
Note that $\bar{\square}\bar{R} = M^2\bar{R}$ is the trace equation of local $R+R^2$ which gives Starobinsky inflationary solution. Here given the conditions \eqref{conditions} are met we can solve exactly the full non-local gravity equations of motion and also we have good inflationary solution. Now it is important to see how non-local effects of gravity appear through inflationary perturbations. Before we go this step we can draw a heuristic bound on the scale of non-locality shown in Fig.~\ref{fig:scales}  based on the fact that we require near scale invariance at the scale of inflation. 
This can be seen 
heuristically by expanding the quadratic Ricci scalar part of the action as 
\begin{equation*}
	S = \int d^4x\sqrt{-g} \LT \frac{M_p^2}{2}R+\frac{M_p^2}{12M^2}R^2+\Oc \LF \frac{M_p^2R\square R}{M^2\Mc_{s}^2}  \RF \RT\,. 
\end{equation*}
In the high curvature regime $R\gg M^2$ local quadratic curvature term is naturally dominant and it is known to be scale invariance. 

\subsection{Scalar and tensor power spectrum}

To study perturbations around quasi-de Sitter we take the perturbed metric of the form 
\begin{equation}
	ds^{2}=a^{2}\left(\eta\right)\left[-\left(1+2\Phi\right)d\eta^{2}+\left(\left(1-2\Psi\right)\delta_{ij}+2h_{ij}\right)dx^{i}dx^{j}\right]\,.\label{line-element}
\end{equation} 
where $\Phi,\,\Psi$ are Bardeen potentials and $h_{ij}$ is tensor fluctuation. 
The study of non-local perturbed equations of motion \cite{Koshelev:2017tvv} has revealed that $\Phi+\Psi\approx 0$ during inflation. Since $\delta W_{\mu\nu\rho\sigma}\propto \LF \Phi+\Psi \RF $ the Weyl tensor part in \eqref{NC-action} does not contribute to the scalar perturbations during inflation. Following the computations in \cite{Koshelev:2017tvv,Koshelev:2020foq} the second order action of scalar degree of freedom in de Sitter approximation becomes 
 \begin{equation}
	\delta^{2}S_{(s)}=\frac{1}{2\Fc_{1}\bar{R}}\int d\tau d^{3}x\sqrt{-\bar{g}}\Upsilon\frac{\Wc\LF\frac{\bar{\square}}{\Mc^{2}}\RF}{\Fc_{R}\LF\frac{\bar{\square}}{\Mc^{2}}\RF}(\bar{\square}-M^{2})\Upsilon\,.
	\label{upsieq}
\end{equation}
where $\Upsilon \approx 2\bar{R}\Psi$. In order not to have any more ghost degrees of freedom we choose the structure of form factor as 
\begin{equation}
		\Fc_{R}\LF \square_s \RF = \Fc_{1}\frac{3e^{\gamma_S\LF \square_s \RF } \LF \square_s-\frac{M^2}{\Mc_{s}^2} \RF+\LF \frac{\bar{ R}}{\Mc_{s}^2}+3\frac{M^2}{\Mc_{s}^2} \RF}{3\square_s+\frac{\bar{ R}}{\Mc_{s}^2}} \,. 
\end{equation}
where $\bar{ R}\approx {\rm constant}$ and $\gamma_S$ is an entire function.  Therefore, the theory now has only one scalar(on) degree of freedom similar to local $R+R^2$ theory but however, the the scalar field here is non-local in nature. 

Computation of the power spectrum and the scalar spectral index of curvature perturbation $\Rc= \Psi+H\frac{\delta R}{\dot{\bar{R}}}=\Psi+\frac{24H^{3}}{24H\dot{H}}\Phi\approx-\frac{1}{\epsilon}\Psi $ we get
\begin{equation}
	\left.\Pc_\Rc= \frac{H^2}{16\pi^2\epsilon^2}\frac{1}{3\lambda\Fc_1\bar{R}} \right|_{k=aH}\,,\quad  n_s\equiv \left.\frac{d\ln \Pc_\Rc}{d\ln k}\right|_{k=aH} \approx 1-\frac{2}{N} \,,\label{Psuni}
\end{equation}

Similarly the computation of the second order action of tensor perturbations gives (with addition of the non-trivial contributions from the Weyl tensor part in \eqref{NC-action})
\begin{equation}
	\begin{aligned}\delta^{2}S_{(T)}= & \frac{1}{2}\int d^{4}x\sqrt{-\bar{g}}h_{ij}^{\perp} e^{2\gamma_T\LF \square_s-\frac{2\bar{ R}}{3\Mc_{s}^2} \RF}\LF\bar{\square}-\frac{\bar{R}}{6}\RF h^{\perp{ij}}\,,\end{aligned}
	\label{tenspart}
\end{equation}
where
\begin{equation}
	\Fc_W\LF \square_s \RF = \frac{\Fc_{1}\bar{ R}}{\Mc_{s}^2}\frac{e^{\gamma_T
			\LF \square_s-\frac{2}{3}\frac{\bar{ R}}{\Mc_{s}^2} \RF}-1}{\square_s-\frac{2\bar{ R}}{3\Mc_{s}^2}}\,.
	\label{fw}
\end{equation}
which fixes the tensor degrees of freedom to be just one massless graviton. 

\begin{figure}
	\centering
	\includegraphics[width=1.1\linewidth]{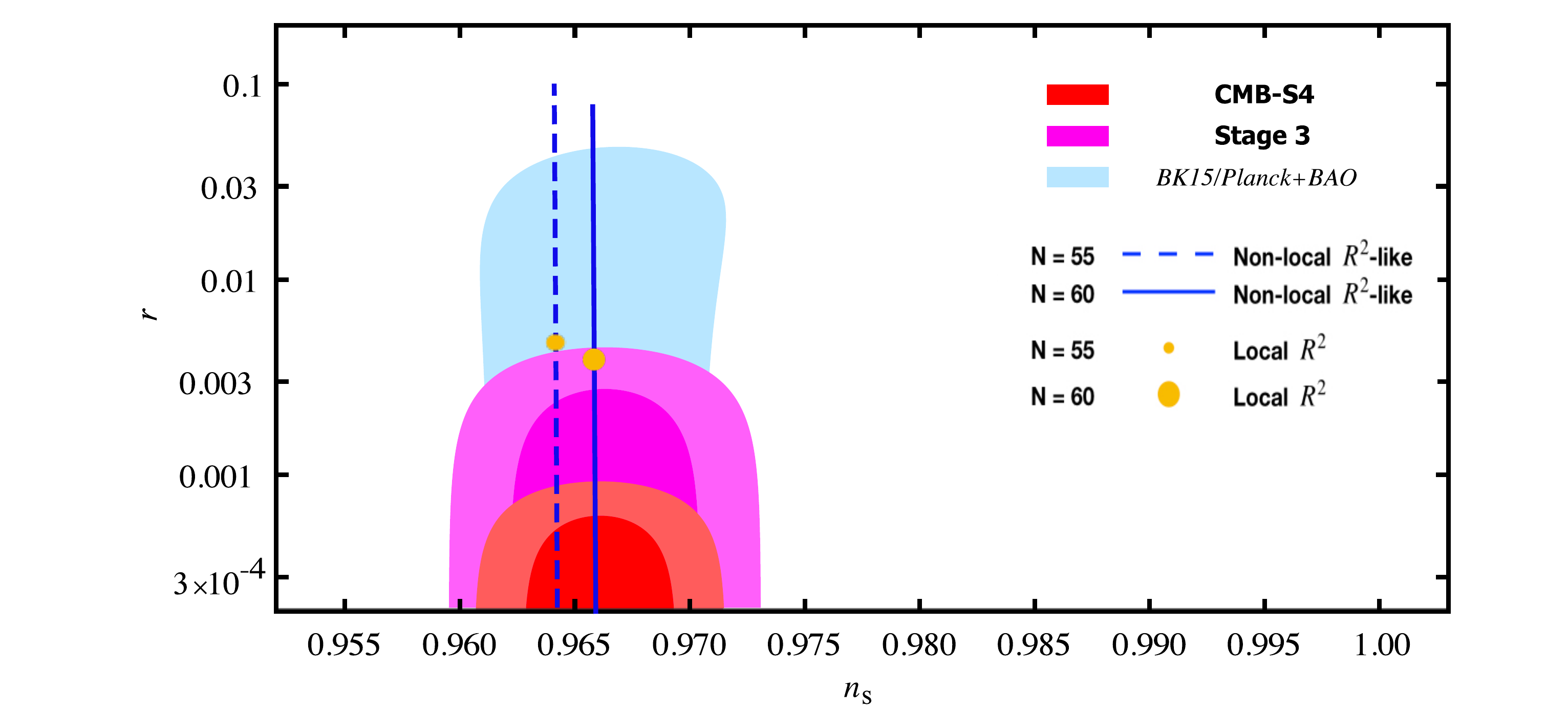}
	\caption{We depict the predictions of non-local $R^2$-like inflation in the $(n_s,r)$ plane of the latest CMB S-4 science paper about future forecast of detecting $B$-modes \cite{Abazajian:2019eic,Koshelev:2017tvv,Koshelev:2020xby}. }
	\label{fig:nsrlm1}
\end{figure}

\begin{figure}[h!]
	\centering
	\includegraphics[width=0.8\linewidth]{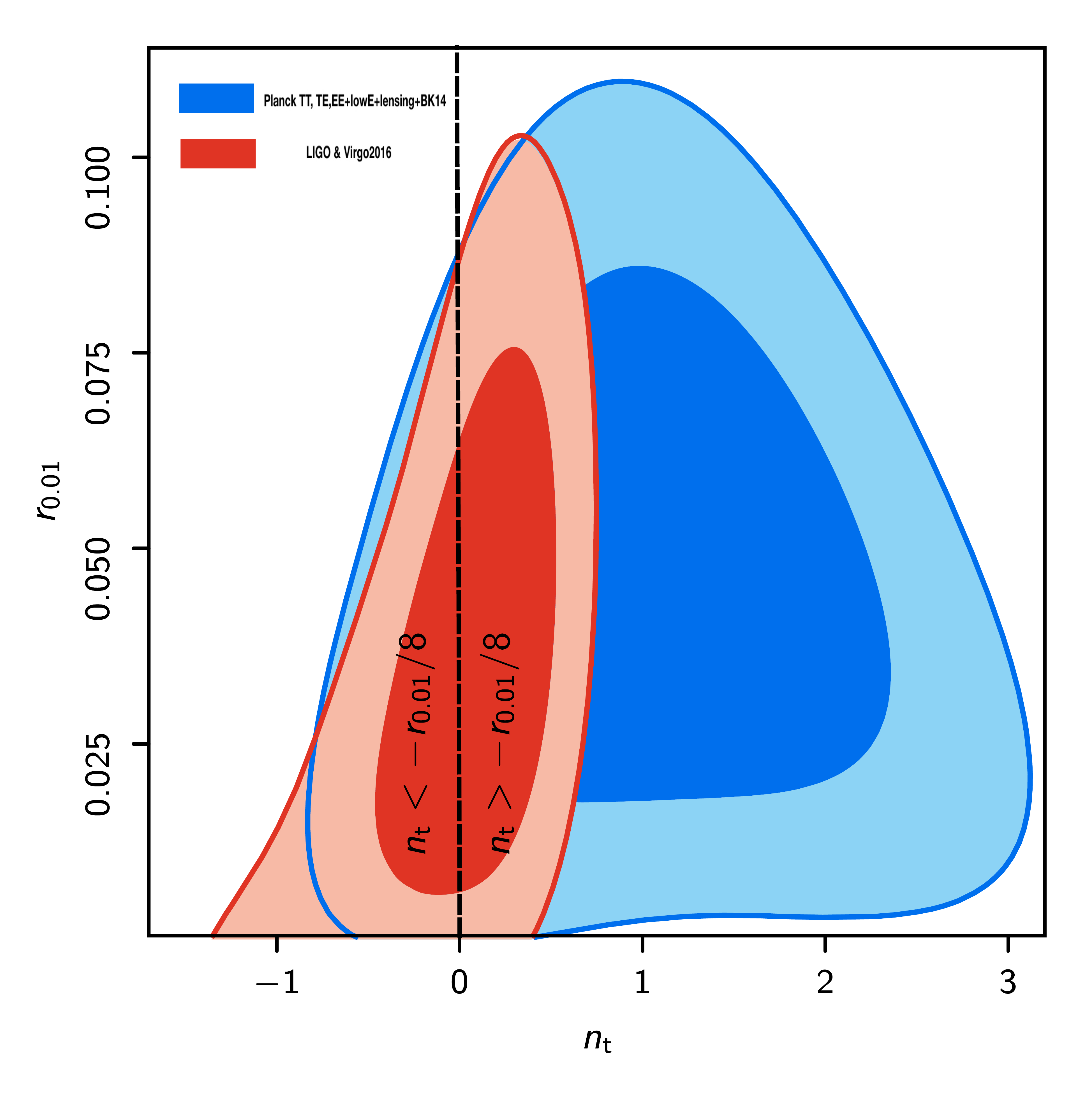}
	\caption{We note that the predictions of non-local $R^2$-like inflation can be anywhere within the likelihood projected $(n_t,r)$ plane of latest Planck 2018  \cite{Abazajian:2016yjj,Koshelev:2017tvv,Koshelev:2020xby}.}
	\label{fig:nt-r}
\end{figure}

Computing the tensor power spectrum in the leading order in slow-roll we obtain 
\begin{equation}
	\left.\mathcal{P}_{\mathcal{T}}\right|_{k=aH}=\frac{H^{2}}{\pi^{2}\mathcal{F}_{1}\bar{R}}e^{-2\gamma_T\left(\frac{\bar{R}}{6\mathcal{M}_s^{2}}\right)}
	\, .\label{hatH-pwt}
\end{equation}
As a result the key  inflationary parameter called tensor-to-scalar ratio becomes 
 \begin{equation}
r= \frac{12}{N^{2}}e^{-2\gamma_T\LF\frac{-\bar{R}}{2\Mc_s^{2}}\RF}\Bigg\vert_{k=aH}
 \end{equation}
The tilt of the tensor power spectrum can be obtained as 
	\begin{equation}
	\begin{aligned}
		n_{t}\equiv\left.\frac{d\ln\mathcal{P}_{\mathcal{T}}}{d\ln k}\right|_{k=aH} & \approx\left.-\frac{d\ln\mathcal{P}_{\mathcal{T}}}{dN}\left(1+\frac{1}{2N}\right)\right|_{k=aH}\\
		& \approx\left.-\frac{3}{2N^{2}}-\left(\frac{2}{N}+\frac{3}{2N^{2}}\right)\frac{\bar{R}}{2\mathcal{M}_s^{2}}\gamma_T^{(\prime)}\left(-\frac{\bar{R}}{2\mathcal{M}_s^{2}}\right)\right|_{k=aH}\,,
	\end{aligned}
	\label{Ttilt}
\end{equation}
where $^\prime$ denotes derivative with respect to the argument. 
The above predictions of non-local $R^2$-like inflation are depicted in Fig.~\ref{fig:nsrlm1} and Fig.~\ref{fig:nt-r}.

\section{Primordial (scalar) non-Gaussianities}

\begin{figure}
	\centering
	\includegraphics[width=0.9\linewidth]{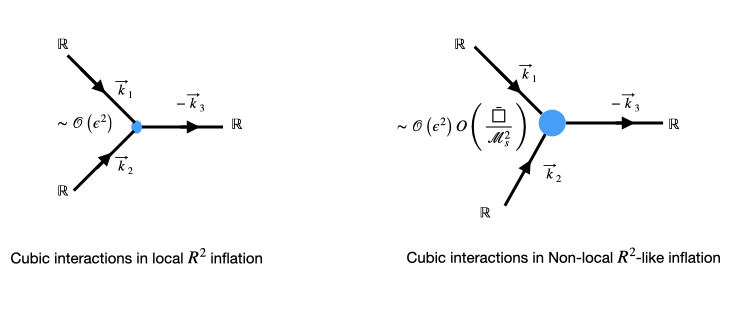}
	\caption{In the above plot $\mathbb{R}= \{ \Rc,\,\partial\Rc,\,\Dc^\mu\partial\Rc \}$ imply various tree level interactions of different modes of the curvature perturbation in the local $R^2$ and the non-local $R^2$-like inflation. $O\LF \frac{\bar{ \square}}{\Mc_{s}^2} \RF$ is some analytic non-local operator.}
	\label{fig:3pts}
\end{figure}

Beyond two point correlations it is interesting to see if there will be non-Gaussianities in non-local $R^2$-like inflation. As we discussed before the scalar(on) in the non-local $R^2$-like inflation is non-local in nature. Therefore, we generally expect the non-local interactions to play a role in predicting inflationary 3-point correlations different from the local theories. The primordial (scalar) non-Gaussianities in the non-local $R^2$-like inflation are computed in \cite{Koshelev:2020foq} and here we briefly review the main results.  The 3-point correlations and the parameter $f_{NL}$ are defined by
\begin{eqnarray*}
	\langle\Rc\left(\mathbf{k_{1}}\right)\Rc\left(\mathbf{k_{2}}\right)\Rc\left(\mathbf{k_{3}}\right)\rangle  =  \left(2\pi\right)^{3}\delta^{3}\left(\mathbf{k_{1}}+\mathbf{k_{2}}+\mathbf{k_{3}}\right)\mathcal{B}_{\Rc}\left(k_{1},k_{2},k_{3}\right)\,,
\end{eqnarray*}
and
\begin{equation*}
	B_{\Rc}\left(k_{1},k_{2},k_{3}\right) = (2\pi)^4\frac{1}{\prod_i k_i^3} \Pc_{\Rc}^{\ast 2} f_{NL}\LF k_1,\,k_2,\,k_3 \RF \,, 
\end{equation*}
where 
\begin{equation}
	f_{NL} = -\frac{5}{6}\frac{A_\Rc\LF k_1,\,k_2,\,k_3 \RF}{\sum_i k_i^3}\,. 
	\label{fnldef}
\end{equation}
The interactions of the curvature perturbations in the non-local $R^2$-like inflation can be schematically explained as in Fig.~\ref{fig:3pts}. 

 To calculate bi-spectrum, first we expand our action to cubic order to the leading order in slow-roll parameter. Then we compute the correlations using the following definition 
 \begin{equation}
 	\langle\Rc\left(\mathbf{k_{1}}\right)\Rc\left(\mathbf{k_{2}}\right)\Rc\left(\mathbf{k_{3}}\right)\rangle  = -i \int_{-\infty}^{\tau_e}  a d\tau \langle 0\vert [ \Rc(\tau_{e},\,\mathbf{k_1})\Rc(\tau_{e},\,\mathbf{k_2})\Rc(\tau_{e},\,\mathbf{k_3}),\, H_{int} ] \vert 0 \rangle \,,
 	\label{3-point-f}
 \end{equation}
 where $\boldsymbol{k}_i$ are wave vectors and $H_{int}$ is the interaction Hamiltonian. It is related to a perturbation of the Lagrangian (\ref{NC-action}) expanded up to the 3rd order in curvature perturbations ($\Lc_3$) as $H_{int}\approx -\Lc_3$. We consider the mode functions as 
 \begin{equation}
 \Rc \approx -\Psi/\epsilon,\quad  \bar{\square}\Rc =M^2\Rc
 \end{equation}
New Interactions arise via the commutation relation in de Sitter approximation
\begin{equation}
	\square \nabla_\mu\Rc = \nabla_\mu\square \Rc +\frac{\bar{ R}}{4} \nabla_\mu\Rc. 	
\end{equation} 
After numerous steps of computation the cubic order action in $\Rc$ of \eqref{NC-action} in the leading order slow-roll approximation is obtained as \cite{Koshelev:2020foq}
		\begin{equation*}
			\begin{aligned}
				\delta^{(3)}S_{(S)} =  & 4\epsilon M_p^2 \int   d\tau d^3x \Bigg\{T_1^\ast \Rc\nabla\Rc\cdot\nabla\Rc+
				T_2^\ast \Rc\Rc^{\prime 2}+ T_3^\ast \Hc^2\Rc^3\\+&
				T_4^\ast 	\Hc \Rc\Rc\Rc^\prime+ T_5^\ast\Hc^{-1}\nabla\Rc\cdot\nabla\Rc\Rc^\prime + T_6^\ast \Hc^{-1}\Rc^{\prime 3} \\ &+ T_7^\ast \Hc^{-2}\Rc^\prime \nabla\Rc\cdot\nabla\Rc^\prime  \Bigg\}\,, 
			\end{aligned}
			\label{3rda}
		\end{equation*}
		where $T_i$'s are dimensionless constants whose can be found in \cite{Koshelev:2020foq}

\begin{figure}
	\centering
	\includegraphics[width=0.7\linewidth]{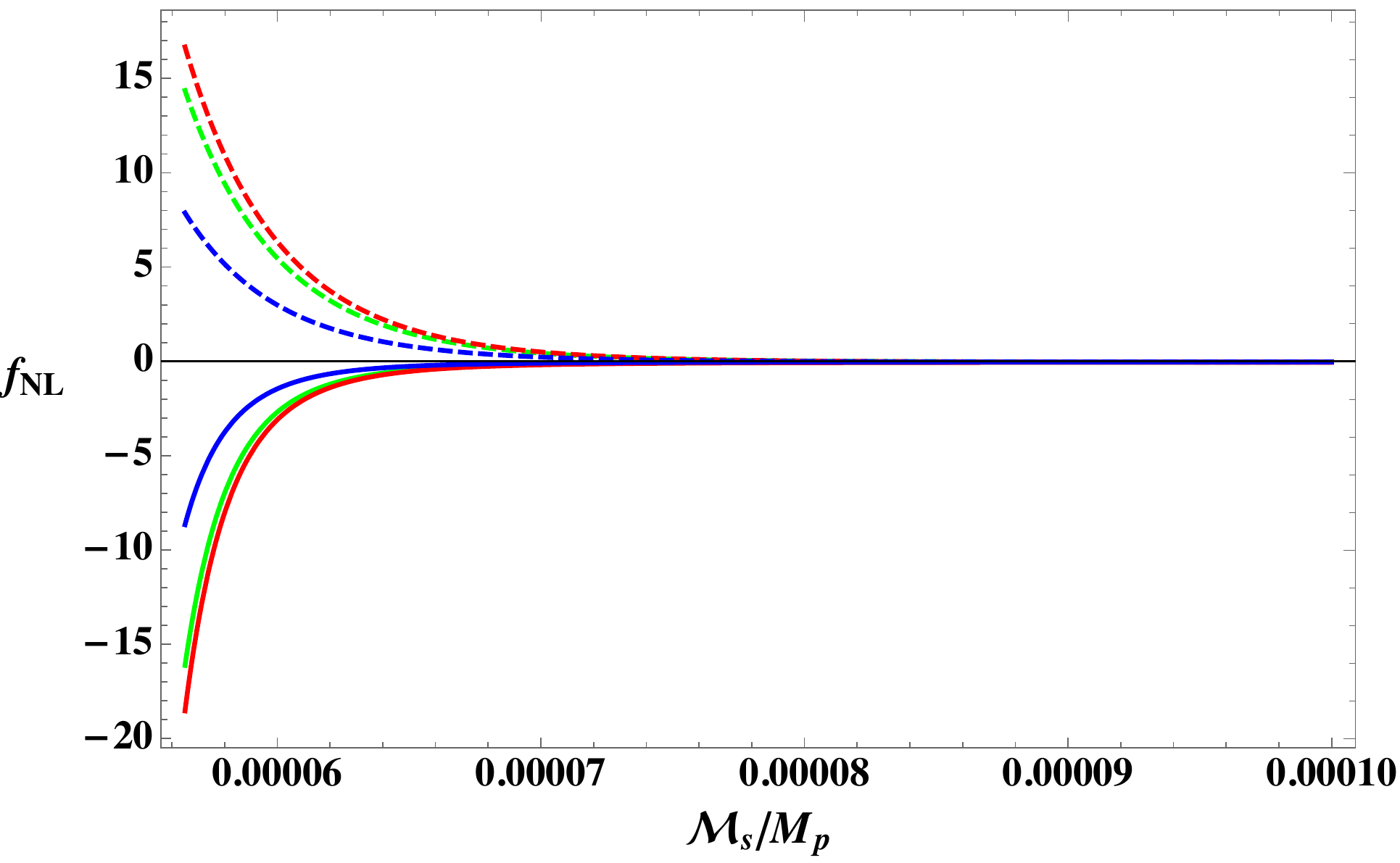}
	\caption{In the above plots, $f_{NL}$ versus the scale of non-locality $\Mc_{s}$ (in the units of $M_p$) is depicted for squeezed (blue), equilateral  (red), and orthogonal (green) configurations for the polynomial entire functions $\gamma_T$ given by eqs. (4.19) and (4.20) of arXiv:2003.00629 and represented by solid and dashed lines respectively. Here $N=55$ of $e$-foldings is assumed. In the limit $\Mc_{s}\to M_p$ the predictions of the local $R^2$ model are recovered.}
	\label{fig:fnl1}
\end{figure}

Computing different shapes of $f_{NL}$ parameter when $k=aH$ we get \cite{Koshelev:2020foq}
\begin{equation}
\begin{aligned}
	f_{NL}^{sq} \Bigg\vert_{k=aH}& \approx  \frac{5}{12} \LF 1-n_s \RF -23 \epsilon^2\left[e^{\gamma_S \left(\frac{\bar{ R}}{4 \Mc_s^2}\right)}-1\right] -\frac{4\bar{ R} }{\Mc_s^2}\epsilon^3 e^{\gamma_S \left(\frac{\bar{ R}}{4 \Mc_s^2}\right)} \gamma _S^{\prime}\left(\frac{\bar{ R}}{4 \Mc_s^2}\right)\,\\ 
	f_{NL}^{eq}  \Bigg\vert_{k=aH} & \approx \frac{5}{12} \LF 1-n_s \RF -49 \epsilon^2\left[e^{\gamma_S \left(\frac{\bar{ R}}{4 \Mc_S^2}\right)}-1\right]-\frac{9 \bar{ R}}{ \Mc_s^2}\epsilon^3 e^{\gamma_S \left(\frac{\bar{ R}}{4 \Mc_s^2}\right)} \gamma _S^{\prime}\left(\frac{\bar{ R}}{4 \Mc_s^2}\right)\\ 
	f_{NL}^{ortho}  \Bigg\vert_{k=aH}& \approx \frac{5}{12} \LF 1-n_s \RF -43 \epsilon^2\left[e^{\gamma_S \left(\frac{\bar{ R}}{4 \Mc_s^2}\right)}-1\right] -\frac{22\bar{ R}_\ast }{3 \Mc_s^2}\epsilon^3 e^{\gamma_S \left(\frac{\bar{ R}}{4 \Mc_s^2}\right)} \gamma _S^{\prime}\left(\frac{\bar{ R}}{4 \Mc_s^2}\right)\,. 
\end{aligned}
\label{fnlf}
\end{equation}
These characterize the non-local interactions of curvature perturbation. Clearly from the above expressions single field Maldacena consistency relation is violated in non-local theories and one can obtain detectable level of non-Gaussianities due to non-local nature of interactions (See Fig.~\ref{fig:fnl1}). This is totally non-trivial effect in comparison with all the known mechanisms of producing large non-Gaussianities in the local theories. 

\section{Concluding remarks} 
We conclude here with highlighting the notable results of early Universe cosmology in the context of UV-non-local ghost-free theories of gravity. First of all we note that the UV non-local gravity action \eqref{NC-action} is a consistent and necessary extension of Stelle gravity in order to achieve UV-completion based on the first principles. Another point to note here is "non-locality" is an important concept that is needed for building a consistent theory of quantum gravity.  We have discussed how one can have a successful realization of inflation with the action action \eqref{NC-action} and how non-local nature of gravity can be seen through inflationary observables. These are the important conclusions we can draw here to probe non-local nature of gravity in the scope of future CMB and gravitational wave observations \cite{Abazajian:2016yjj,Abazajian:2019eic,Meerburg:2016zdz,Meerburg:2019qqi,Ricciardone:2016ddg}. 
\begin{itemize}
	\item In the non-local $R^2$-like inflation the primordial scalar power spectrum remains same as in the case of local $R+R^2$ inflation. However the tensor power spectrum is modified and one can have any value of tensor-to-scalar ratio\cite{BICEP:2021xfz,Koshelev:2020foq,Koshelev:2020xby} $r<0.036$. As a result of modification of tensor power spectrum the tensor tilt in this theory is modified and ultimately tensor consistency relation $r=-8n_t$ is violated. Note that the violation of consistency relation is purely a non-local effect, the sound speed of primordial fluctuations are unity in this theory. 
	\item The blue tensor tilt $n_t>0$ is possible with non-local $R^2$-like inflation. Since not many frameworks of inflation can predict blue tilt any detection of it automatically favours non-local theory. 
	Also $n_t<0$ can also equally possible too. 
	\item Non-local nature of gravity leads to detectable range of non-Gaussianities especial one can have squeezed limit of $f^{\rm eq}_{NL}\sim O(1)$ despite  there is only one scalar degree of freedom and nature of inflation is standard-slow-roll. Further more, other shapes of $f_{NL}$ can also be large despite the sound speed of primordial scalar field is unity. 
\end{itemize}

In summary, non-local nature of gravity brings very unique predictions which are indeed very interesting in the view of future observations as well as further theoretical investigations. 

\section{Acknowledgments}
KSK acknowledges financial
support from JSPS and KAKENHI Grant-in-Aid for Scientific Research No. JP20F20320. 

\bibliographystyle{utphys}
\bibliography{ssa}

\end{document}